\documentclass[11pt]{article}
\usepackage[T1]{fontenc}
\usepackage[utf8]{inputenc}
\usepackage{lmodern}
\usepackage[margin=1in]{geometry}
\usepackage{booktabs}
\usepackage{amsmath,amssymb}
\usepackage{graphicx}
\usepackage{xurl}
\usepackage{hyperref}
\usepackage{microtype}
\usepackage{array}
\usepackage{natbib}
\title{Assessing the Reliability of Longitudinal Bibliometric Rankings under Incomplete Citation Histories}
\author{Jo\~ao Paulo Papa\\\small School of Sciences, S\~ao Paulo State University}
\date{November 12, 2026}
\begin{document}
\maketitle
\begin{abstract}
Longitudinal bibliometric analyses rely on annual citation histories, yet missing annual observations are frequently difficult to distinguish from true zero-citation years. We present a reproducible case-study framework for auditing citation-history gaps, tracing retrieval provenance, and quantifying the sensitivity of longitudinal indicators and rankings to explicit missing-data assumptions. The case study comprises 595 consolidated publications derived from 616 Google Scholar article records, and examines the completed calendar years 2024--2025. Three scenarios are compared: strictly observed annual counts (A), hypothetical zeros for internal gaps only (B), and hypothetical zeros for all missing years within a valid publication interval (C). The number of publications with computable growth and acceleration indicators increases from 196 (A) to 257 (B) and 490 (C); the corresponding counts for a revival indicator are 131, 152, and 177. Although the ordering of publications eligible under both compared scenarios is unchanged (Spearman and Kendall correlations equal to 1.0), complete-ranking positions and Top-$K$ membership may change markedly. In particular, acceleration exhibits a Jaccard overlap of 0.563 at $K=50$ between A and C, while revival exhibits 0.667 at $K=10$. These findings demonstrate that unchanged indicator values and perfect correlations on a common subset do not guarantee stable rankings when eligibility changes. The zero assignments are hypothetical sensitivity assumptions, not evidence of provider semantics. The results are descriptive and based on a single researcher profile. A reusable Google Scholar Tracker software workflow supports the audit and sensitivity analysis; version 1.8.5 is publicly available on GitHub and archived on Zenodo (DOI: 10.5281/zenodo.24170285).
\end{abstract}
\noindent\textbf{Keywords:} bibliometrics; citation history; missing data; ranking robustness; sensitivity analysis; reproducibility.

\section{Introduction}
\label{sec:introduction}

Quantitative indicators have become an established component of the analysis and evaluation of scientific activity. Citation-based measures are used to describe the visibility of publications, characterize research portfolios, and complement expert assessment. Yet a citation indicator is not an intrinsic property of a publication: its interpretation depends on the underlying bibliographic database, the definition of the counted items, the observation period, and the computational procedure. The Leiden Manifesto emphasizes transparency, scrutiny of data and indicators, and the need to account for the limitations of quantitative research assessment \citep{hicks2015leiden}. Likewise, comprehensive reviews of citation impact indicators highlight the methodological consequences of database selection, publication coverage, and counting conventions \citep{waltman2016review}. These concerns are especially consequential when the intended analysis goes beyond cumulative citation totals.

A cumulative citation count summarizes the attention received by a publication up to a particular collection date, but it does not reveal when that attention occurred. Two publications with similar cumulative totals may exhibit markedly different temporal trajectories: one may have accumulated citations soon after publication and subsequently declined, whereas another may be attracting increasing attention or experiencing renewed interest years later. Annual citation histories make it possible to examine such patterns through measures of recent citation volume, changes in annual citation rates, and the re-emergence of older publications. In contrast to a cumulative snapshot, however, a longitudinal indicator generally requires observations for a specific set of calendar years. The validity of the resulting measure therefore depends not only on the observed citation counts but also on the meaning of years for which no count is available.

This distinction creates a fundamental data-quality problem. An explicitly recorded annual citation count of zero is an observation. An absent annual record is not, by itself, evidence of zero citations. Depending on the provider, its interface, and the collection pipeline, an omitted year could reflect a zero-suppression convention, incomplete coverage, a retrieval failure, a parsing decision, or a historical revision. These explanations have different implications, yet they may appear identical in a downstream dataset that stores only returned year--count pairs. Replacing every missing year with zero can therefore introduce an unverified assumption; conversely, requiring every annual value to be explicitly observed can substantially reduce the number of publications eligible for longitudinal analysis. This trade-off is particularly important when the purpose is to compare or rank publications rather than merely display their histories.

Google Scholar provides a useful setting in which to examine this problem. Its role as an accessible source of citation information has been discussed alongside questions of coverage, source heterogeneity, and bibliographic-data quality \citep{harzing2008google}. Nevertheless, a Google Scholar profile, the annual histories exposed through a retrieval service, and a locally maintained longitudinal database are not interchangeable representations. The profile may contain multiple records corresponding to one publication, annual observations may be available for only a subset of records, and cumulative snapshots may have been collected at different times from annual series. A defensible analysis must therefore retain the provenance of each observation and distinguish publication-level consolidation from historical-data reconstruction. In particular, annual histories associated with different bibliographic records should not be combined automatically merely because those records are believed to represent the same work.

The implications of missing observations extend beyond the numerical values of indicators. Consider a ranking based on the difference between citations received in two consecutive years. If both annual counts are observed, the difference is directly computable. If one count is absent, the publication is ineligible under a conservative policy. Assigning a hypothetical zero may make the indicator computable and allow the publication to enter the ranking. Even when the values of all previously eligible publications remain unchanged, newly eligible publications can displace them in the full ranking or alter membership in a Top-$K$ list. Thus, a perfect rank correlation calculated on the publications shared by two scenarios need not imply that the complete rankings are stable. This separation between \emph{indicator-value stability}, \emph{eligibility stability}, and \emph{ranking-membership stability} motivates the present study.

We investigate the following overarching research question: \emph{To what extent do alternative, explicitly stated interpretations of missing annual citation observations affect the computability and robustness of longitudinal bibliometric rankings?} We address it through three more specific questions:
\begin{enumerate}
\item \textbf{RQ1---Coverage and provenance:} What kinds of gaps occur in the annual citation histories of a consolidated publication portfolio, and which uncertainties can be resolved from stored observations and retrieval metadata?
\item \textbf{RQ2---Eligibility sensitivity:} How does the number of publications with computable longitudinal indicators change when internal gaps, or all valid missing years, are treated as hypothetical zeros?
\item \textbf{RQ3---Ranking robustness:} How do these assumptions affect Top-$K$ membership, relative ordering within a common eligible universe, and absolute positions in the complete rankings?
\end{enumerate}

To answer these questions, we develop and evaluate a provenance-aware, reproducible workflow using one Google Scholar researcher profile as a methodological case study. The source dataset comprises 616 article records represented by 595 consolidated publications. The principal analysis concerns the completed calendar years 2024 and 2025. We distinguish leading, internal, and trailing gaps, as well as publications without usable annual history; we also track uncertain publication dates and limitations of the available retrieval evidence. Three scenarios are evaluated without altering the stored data: (A) observed annual counts only, (B) hypothetical zeros restricted to internal gaps, and (C) hypothetical zeros for all missing years in a valid publication interval. We apply the existing growth, acceleration, and revival indicators consistently across these scenarios and assess both indicator computability and ranking stability.

The case study provides a concrete illustration of why these dimensions must be reported separately. Under the observed-only scenario, growth and acceleration are computable for 196 consolidated publications; the number rises to 257 when internal gaps are hypothetically set to zero and to 490 under the most permissive scenario. At the same time, the relative ordering of publications eligible in both compared scenarios remains unchanged, whereas Top-$K$ composition is less stable for certain indicators. For example, the Jaccard overlap between the conservative and most permissive acceleration Top-50 lists is approximately 0.563. These figures are not estimates of the true prevalence of zero-citation years. Rather, they quantify the consequences of competing assumptions whose validity is not established by the retained provider evidence.

The main contributions of this work are threefold. First, we present an explicit gap taxonomy and a provenance-aware procedure that separates observed zero values, missing annual records, unsuccessful retrievals, and publication-year constraints. Second, we provide a non-destructive sensitivity protocol that measures changes in indicator eligibility without silently imputing historical observations. Third, we introduce a complementary ranking analysis that distinguishes Top-$K$ overlap and complete-ranking displacement from correlations computed only among publications eligible under both scenarios. Fourth, we document a reusable Google Scholar Tracker software workflow that implements the audit, scenario comparisons, and machine-readable reporting; its public release remains planned. The framework is intended to support transparent longitudinal bibliometric analysis, not to establish a new measure of scientific quality or to generalize from a single researcher profile to the entire scholarly population.

The remainder of the paper is organized as follows. Section~\ref{sec:related} positions the study in the literature on bibliometric indicators and data quality. The subsequent sections describe the data model and missingness scenarios, present the experimental protocol, report eligibility and ranking-robustness results, discuss their implications and limitations, and conclude with directions for independent replication and prospective provider-response validation.

\section{Related Work and Positioning}
\label{sec:related}
\subsection{Citation indicators and responsible interpretation}
Citation indicators are widely used to describe research visibility, but their interpretation is conditional on the database, the citation window, the unit of analysis, and the counting method. Waltman's review organizes the methodological literature around bibliographic sources, publication selection, normalization, and counting conventions \citep{waltman2016review}. The Leiden Manifesto further argues for transparent data and analytical procedures, contextual interpretation, and scrutiny of indicators used in research evaluation \citep{hicks2015leiden}. These principles are especially relevant for longitudinal indicators, which require more than a single cumulative citation count. Our study does not propose a measure of research quality or a replacement for expert assessment. Instead, it investigates whether the available temporal evidence is sufficient to compute and compare selected citation-trajectory descriptors.

\subsection{Bibliographic coverage and source-dependent measurement}
Google Scholar has attracted attention because its indexing extends beyond the journal-focused coverage of traditional citation databases. \citet{harzing2008google} discuss both its potential value for citation analysis and practical limitations associated with source heterogeneity and bibliographic accuracy. In a systematic comparison across 252 subject categories, \citet{martin2018systematic} report that Google Scholar identifies a substantial number of citations not indexed by Web of Science or Scopus, including citations from non-journal materials. Broader coverage, however, should not be equated with uniformly reliable or complete longitudinal histories. Cross-database comparisons primarily characterize the coverage of publications and citing sources; they do not, by themselves, establish the meaning of an absent annual observation in a specific provider response.

This distinction is central to the present work. We analyze annual citation counts returned through a retrieval service for one Google Scholar profile, rather than asserting that the same missingness behavior holds across all citation databases or retrieval interfaces. A publication may have a cumulative citation total even when its locally stored annual history lacks one or more years. We therefore avoid reconstructing annual series from cumulative totals, which may reflect different collection dates or coverage conditions.

\subsection{Record identity, consolidation, and provenance}
Bibliometric analyses require a well-defined unit of analysis. Source records can represent different versions of one scholarly work or contain incomplete or inconsistent bibliographic metadata; the resulting ambiguity matters when computing publication-level indicators. As emphasized in methodological reviews, the selection and counting of publications and citations are substantive choices rather than neutral implementation details \citep{waltman2016review}. For longitudinal analysis, consolidation introduces an additional problem: different source records associated with one work may expose different annual histories. Simply summing or splicing these histories can create observations that were never supplied for the consolidated publication.

Our workflow consequently distinguishes the original source records, reviewed publication identities, and the selected historical reference for each consolidated publication. Duplicate review and retrieval provenance are retained as auditable analytical decisions. The current case study does not establish a generally optimal entity-resolution algorithm; its contribution is to make consolidation assumptions explicit and prevent the accidental mixing of incompatible annual observations.

\subsection{Incomplete temporal observations and sensitivity assumptions}
The methodological distinction between an observed value and a missing observation is particularly consequential for temporal bibliometric data. An explicitly stored zero represents a recorded annual count; an absent year may instead reflect the provider's presentation rules, an incomplete response, a failed retrieval, or downstream parsing. Without evidence of the provider's omission semantics, zero-filling is an assumption rather than a factual correction. Conversely, complete-case requirements can reduce the analyzable portfolio and potentially change which publications are represented in a ranking.

Rather than infer a statistical missingness mechanism from unavailable evidence, we conduct a deterministic sensitivity analysis. The conservative scenario uses only observed annual counts; two alternatives treat internal gaps or all valid missing years as hypothetical zeros. These scenarios are not estimates of missing citations and should not be interpreted as a multiple-imputation procedure. They are controlled counterfactual assumptions that expose how much of the analysis depends on unverified data semantics. Gap categories and publication-year constraints make those assumptions inspectable.

\subsection{Longitudinal rankings and robustness under changing eligibility}
Annual citation indicators can capture recent activity or changes in citation rate that cumulative totals conceal. Yet a ranking built from these indicators is defined jointly by its scores and its eligible population. Two scenarios may preserve every score among previously eligible publications while admitting new publications with potentially high or low scores. Consequently, correlation computed after restricting both rankings to the shared eligible universe answers a different question from Top-$K$ overlap or changes in complete-ranking positions.

We operationalize this distinction through complementary diagnostics: eligibility counts, numerical score stability, Spearman and Kendall correlations on the common eligible universe, Jaccard overlap of Top-$K$ sets, and displacement within the full scenario-specific rankings. Ties are handled explicitly because discrete annual counts can produce many identical scores, particularly for differences between consecutive years. A deterministic display order ensures reproducibility but does not eliminate the need to disclose Top-$K$ boundary ties.

\subsection{Research gap and positioning}
The cited literature provides essential foundations for responsible indicator interpretation, comparative database coverage, and methodological transparency \citep{hicks2015leiden,waltman2016review,harzing2008google,martin2018systematic}. Our narrower question concerns the downstream consequences of incomplete annual citation histories for \emph{both} indicator computability and ranking membership. We do not claim that missing-data sensitivity or ranking comparison is novel in isolation. The proposed contribution is an integrated, provenance-aware case-study protocol that separates observed zeros from absent records, preserves consolidation decisions, and evaluates eligibility, common-universe ordering, and complete-ranking composition under explicit, non-persistent assumptions. The evidence supports a reproducible demonstration on one profile; cross-provider and cross-disciplinary generality remain questions for future studies.

\section{Framework and Methods}
\label{sec:methods}

\subsection{Overview and units of analysis}
\label{sec:framework}
The proposed workflow comprises five logically separated stages: (i) acquisition and preservation of bibliographic records and citation snapshots; (ii) review of potentially duplicated records and construction of consolidated publication identities; (iii) selection and auditing of annual citation histories; (iv) computation of longitudinal indicators under explicit missing-data scenarios; and (v) comparison of eligibility, scores, and ranking stability. Importantly, stages (iii)--(v) operate on a fixed local dataset. Counterfactual observations are generated in memory and are never persisted as measured citation counts.

Let $\mathcal{B}$ denote the set of source bibliographic records and $\mathcal{P}$ the set of consolidated publications. A mapping $\pi:\mathcal{B}\rightarrow\mathcal{P}$ associates each record with its reviewed publication identity. Consolidation is a bibliographic operation, not a rule for adding citation histories: each publication $i\in\mathcal{P}$ has a documented reference record $h(i)\in\pi^{-1}(i)$ from which the historical series is read. The histories of distinct members of $\pi^{-1}(i)$ are not summed or spliced. This restriction prevents a consolidated series from containing values that were never jointly observed for a single source record.

\subsection{Observation model and provenance}
For publication $i$ and calendar year $t$, define $c_{i,t}\in\mathbb{Z}_{\geq 0}$ only when an annual citation count was explicitly stored. Let $O_i$ be the set of years with such observations, and define an observation indicator
\begin{equation}
 o_{i,t}=\begin{cases}1,&t\in O_i,\\0,&t\notin O_i.\end{cases}
\end{equation}
Thus, $o_{i,t}=0$ does \emph{not} imply $c_{i,t}=0$: the latter is undefined without an observation or a stated counterfactual assumption. Annual counts are calendar-year counts, not cumulative totals. A cumulative citation snapshot is retained as a separate object with its own retrieval time and must not be used to reconstruct missing annual values by subtraction.

The provenance record distinguishes the selected historical reference, stored year--count pairs, collection and refresh outcomes where available, and relevant uncertainty flags. A successful request with no usable history is different from a failed request; neither outcome establishes that citations were zero. Since complete historical provider responses were not systematically archived, the present dataset cannot establish whether omitted rows correspond to zero suppression, incomplete provider coverage, or other causes. Our analysis deliberately preserves this uncertainty.

\subsection{Valid temporal domain and gap taxonomy}
Let $p_i$ be a reliable publication year when available, and let $W$ be the set of calendar years needed for a particular indicator. The valid temporal domain $V_i$ includes only years consistent with the reliable publication interval and the analysis horizon. Years preceding $p_i$ are not treated as missing eligible observations; uncertain publication years are flagged rather than silently resolved. Observations that apparently precede the recorded publication year are retained for audit, not automatically corrected, because online-first publication and bibliographic metadata may differ.

Within the valid temporal domain, missing years are classified using the observed history $O_i$. An \emph{internal gap} is a missing year strictly between the earliest and latest observed years; a \emph{leading gap} precedes the first observed year but lies within the valid interval; and a \emph{trailing gap} follows the last observed year. A publication with no usable observations is classified separately as having \emph{no history}. These classes describe the geometry of stored observations rather than the underlying missingness mechanism. Publication-level flags can coexist: for example, a series may have both internal and trailing gaps.

For a year $t$ in the valid domain, define the internal-gap predicate
\begin{equation}
 g^{\mathrm{int}}_{i,t}=\mathbf{1}\!\left[t\notin O_i\ \land\ \exists u,v\in O_i:\ u<t<v\right].
\end{equation}
This definition avoids treating a missing year outside the observed span as an internal gap. All gap counts in the audit refer to the chosen consolidated history, not a union of histories from duplicate records.

\subsection{Three counterfactual missing-data scenarios}
\label{sec:scenarios}
We consider scenarios $s\in\{A,B,C\}$, where $A$ is the conservative reference and $B$ and $C$ are progressively more permissive \emph{sensitivity assumptions}. For each publication-year pair define an availability predicate $a^{(s)}_{i,t}$ and scenario-specific value $\widetilde c^{(s)}_{i,t}$:
\begin{align}
 a^{(A)}_{i,t}&=o_{i,t},\\
 a^{(B)}_{i,t}&=o_{i,t}\lor g^{\mathrm{int}}_{i,t},\\
 a^{(C)}_{i,t}&=o_{i,t}\lor\mathbf{1}[t\in V_i].
\end{align}
Whenever $a^{(s)}_{i,t}=1$, the value is
\begin{equation}
 \widetilde c^{(s)}_{i,t}=\begin{cases}
 c_{i,t},&o_{i,t}=1,\\
 0,&o_{i,t}=0\text{ and }a^{(s)}_{i,t}=1.
 \end{cases}
\end{equation}
Otherwise the value remains undefined. Scenario $B$ therefore fills only internal gaps, while scenario $C$ hypothetically fills every missing year permitted by the reliable publication interval and the indicator's required years. The assumptions are nested with respect to availability, $a^{(A)}_{i,t}\leq a^{(B)}_{i,t}\leq a^{(C)}_{i,t}$, wherever the same temporal domain applies. These are deterministic \emph{what-if} scenarios, not statistically estimated imputations, and neither $B$ nor $C$ asserts the provider's actual treatment of omitted years.

\subsection{Indicator computability and longitudinal scores}
\label{sec:indicators}
For indicator $m$, let $Q_{i,m}$ be its required set of calendar years, together with any publication-age prerequisites. A publication is eligible under scenario $s$ if the relevant age and metadata constraints hold and every required annual count is available:
\begin{equation}
 e^{(s)}_{i,m}=\mathbf{1}[\text{metadata and age conditions hold}]\prod_{t\in Q_{i,m}}a^{(s)}_{i,t}.
\end{equation}
The eligible universe is $E_{m,s}=\{i\in\mathcal P:e^{(s)}_{i,m}=1\}$. Computability is distinct from satisfying a positive-growth or revival-detection criterion. The reported rankings include all computable publications, including those with non-positive acceleration.

For the completed-year analysis window $W=\{2024,2025\}$, the implemented \emph{growth} indicator is recent citation volume, not a percentage growth rate:
\begin{equation}
 G^{(s)}_i=\widetilde c^{(s)}_{i,2024}+\widetilde c^{(s)}_{i,2025}.
\end{equation}
The \emph{acceleration} indicator is the difference between consecutive annual citation counts:
\begin{equation}
 A^{(s)}_i=\widetilde c^{(s)}_{i,2025}-\widetilde c^{(s)}_{i,2024}.
\end{equation}
A positive acceleration is flagged separately. The existing \emph{revival} indicator evaluates age-eligible publications (at least five years old) using percentage growth in the latest annual count relative to the mean of the preceding three years. Its pre-existing detection criterion requires the latest count to exceed that mean and to meet a 50\% growth threshold. For exact reproducibility, the implementation's convention for a zero preceding mean, percentage scaling, and any boundary cases must be transcribed from the tested source code before submission; the current experiment report does not specify those details. We do not substitute a guessed formula. The corresponding eligibility and ranking results are taken directly from the tested v1.8.5 outputs.

\subsection{Scenario-specific ranking and ties}
\label{sec:rankings}
For each indicator $m$ and scenario $s$, let $f_{m,s}(i)$ denote its score for $i\in E_{m,s}$. The full ranking $R_{m,s}$ orders eligible publications by descending score. Because annual counts are discrete, distinct publications can share a score. The implementation reports competition ranks and applies a deterministic display order using the latest annual citations where applicable, then title and consolidated publication identifier. These secondary keys make the Top-$K$ set reproducible but do not make a score tie scientifically meaningful. Boundary ties are therefore diagnosed and must be considered when interpreting membership changes.

For $K\in\{10,20,50\}$, define $T^{(K)}_{m,s}$ as the first $\min(K,|E_{m,s}|)$ publications in the deterministic display order. For two scenarios $s$ and $s'$, the Jaccard overlap and overlap coefficient are
\begin{align}
 J^{(K)}_m(s,s')&=\frac{|T^{(K)}_{m,s}\cap T^{(K)}_{m,s'}|}{|T^{(K)}_{m,s}\cup T^{(K)}_{m,s'}|},\\
 \Omega^{(K)}_m(s,s')&=\frac{|T^{(K)}_{m,s}\cap T^{(K)}_{m,s'}|}{\min(|T^{(K)}_{m,s}|,|T^{(K)}_{m,s'}|)}.
\end{align}
Empty-set cases are handled explicitly by the implementation. We also enumerate entrants and exits by consolidated identifier and distinguish changes in eligibility from changes in Top-$K$ membership.

\subsection{Shared-universe ordering and full-ranking displacement}
\label{sec:stability}
A central methodological distinction concerns the universe over which rankings are compared. Define the common eligible set $U_m(s,s')=E_{m,s}\cap E_{m,s'}$. To assess whether the relative ordering of previously comparable publications changes, both scenario rankings are restricted to $U_m(s,s')$ and ranks are recomputed \emph{within this fixed set}. Spearman's $\rho$ and Kendall's $\tau_b$ are calculated on these shared-universe scores, with average ranks for ties and appropriate handling of undefined statistics. Neither ineligible publications nor artificial bottom ranks are inserted into this calculation.

Separately, let $r_{m,s}(i)$ be the reported rank of $i$ in the \emph{complete} scenario-specific ranking. Its full-universe displacement is
\begin{equation}
 \Delta r_{m,i}(s,s')=r_{m,s'}(i)-r_{m,s}(i),\qquad i\in U_m(s,s').
\end{equation}
We summarize the mean, median, and maximum of $|\Delta r_{m,i}|$ across the common universe. These values can be large even when $\rho=\tau_b=1$: adding newly eligible publications can shift the original publications' positions without changing their scores or relative ordering. Furthermore, competition ranks and deterministic display positions are not interchangeable when scores tie; the report's rank convention must accompany every displacement statistic.

\subsection{Interpretation and reproducibility safeguards}
We interpret the analysis along four separate dimensions: (1) \emph{coverage}, measured by $|E_{m,s}|$; (2) \emph{score stability} for publications eligible in both scenarios; (3) \emph{relative-order stability} within $U_m(s,s')$; and (4) \emph{composition and positional stability} in complete rankings. No single statistic represents all four dimensions. In particular, a perfect common-universe correlation is compatible with substantial Top-$K$ turnover or full-ranking displacement.

The workflow is designed for read-only execution over a fixed snapshot. It does not query the external citation service, alter annual observations, revise duplicate decisions, or write hypothetical zeros to persistent storage. Versioned JSON, CSV, and LaTeX exports expose scenario definitions, publication identifiers, eligibility counts, ties, ranking metrics, and analysis parameters. The experiments are descriptive: sensitivity scenarios reveal dependence on assumptions but cannot establish which missing observations truly correspond to zero citations.

\section{Experimental Protocol}
\label{sec:protocol}

\subsection{Research design and unit of analysis}
We conduct a retrospective, deterministic case study using the locally stored data of one Google Scholar researcher profile. The experimental unit is a \emph{consolidated publication}, rather than an individual source record, citation event, or researcher. The dataset contains 616 source article records and 595 consolidated publications after reviewed duplicate resolution. The available profile-level information includes two cumulative-citation snapshots; the later snapshot reports 18,814 citations. These totals contextualize the portfolio but are not used to infer missing annual counts or to calculate the indicators evaluated here. The project contains 2,594 article-level historical annual observations; this number is not the number of distinct consolidated publication--year pairs because source-record and consolidated-publication units differ.

All experiments use the same frozen local dataset and the same consolidation decisions. For each consolidated publication, the workflow selects its existing historical reference; it does not concatenate or sum annual histories across candidate duplicate records. This fixed-data design isolates the effect of alternative missing-observation assumptions from changes in data acquisition, record linkage, and snapshot timing. Consequently, the study is a within-dataset sensitivity analysis, not a comparison of independent bibliographic samples.

\subsection{Temporal scope and indicator-specific eligibility}
The principal evaluation window is $W=\{2024,2025\}$, comprising two completed calendar years at the November 2026 analysis date. The incomplete 2026 calendar year is excluded to avoid comparing a partial annual count with completed years. The implementation uses the existing definitions of Growth, Acceleration, and Revival described in Section~\ref{sec:methods}; their required historical observations and metadata conditions determine eligibility independently for each indicator. In particular, Revival has an age requirement and uses a historical baseline, so its eligible set need not equal the eligible sets for Growth and Acceleration.

An observed annual zero is retained as a measured value, while an absent year remains unknown under the primary scenario. No citation counts are inferred from cumulative totals, and years before a reliable publication year are not converted into hypothetical zero observations. Publication-year uncertainty and apparently pre-publication observations are flagged for interpretation rather than silently repaired. The analysis does not presume that a successful retrieval with an empty history indicates zero citations.

\subsection{Audit of historical coverage and chronology}
Before ranking, the framework characterizes the stored historical series according to the gap taxonomy of Section~\ref{sec:methods}. The audit reports 246 consolidated publications with complete applicable requested-year coverage, 61 with internal gaps, 38 with leading gaps, 150 with trailing gaps, and 135 without usable history. It additionally flags 34 publications with uncertain publication year. These are diagnostic categories, not mutually exclusive partitions; their counts must therefore not be summed to recover the portfolio size.

At the observation level, the audit reports 73 internal, 39 leading, 265 trailing, and 188 no-history missing-year instances under its own applicable-year rules. These descriptive counts must not be equated directly with the number of hypothetical zeros used in any particular indicator: the scenario engine also applies publication-year validity and indicator-specific requirements. The chronology audit reports 38 \emph{possible inconsistency checks}, rather than 38 necessarily distinct erroneous publications. Potential explanations include differences between online-first and bibliographic publication dates. The audit cannot resolve such cases without additional primary-source verification.

\subsection{Missing-data sensitivity conditions}
We evaluate three pre-specified scenarios on the identical stored observations: $A$, strictly observed-only; $B$, internal missing years treated as hypothetical zeros; and $C$, all missing years within the valid publication interval treated as hypothetical zeros. The scenarios are nested in terms of permissible hypothetical observations, but none is asserted to represent the actual citation history. The conservative scenario $A$ serves as the reference for primary interpretation. Scenarios $B$ and $C$ test how strongly indicator availability and ranking composition depend on increasingly permissive zero assumptions.

For every indicator $m$ and scenario $s$, we compute the eligible set $E_{m,s}$ and the corresponding indicator scores. The eligibility analysis reports $|E_{m,s}|$ and changes relative to $A$. For publications in the intersection $E_{m,s}\cap E_{m,s'}$, we separately inspect whether the numerical score changes. This distinction is essential: the addition of previously ineligible publications can alter ranking membership even if all scores in the original eligible population are unchanged.

\subsection{Ranking construction and comparison plan}
We rank all computable publications for each indicator--scenario pair, including publications that do not meet an optional positive-trend or revival-detection threshold when the underlying indicator remains computable. Thus, \emph{computability} is not confused with \emph{positive detection}. Rankings use the project's documented competition-rank convention by indicator value, with deterministic secondary ordering for display; correlations on the shared universe use average ranks for ties. We evaluate $K\in\{10,20,50\}$ and the three scenario pairs $A$--$B$, $A$--$C$, and $B$--$C$.

For each pair and indicator, we report the number of eligible publications, the shared eligible-universe size, and the overlap of the complete scenario-specific Top-$K$ sets. Jaccard overlap is the principal membership statistic; the overlap coefficient provides an additional normalization. To assess relative ordering without confounding changes in eligibility, we recompute ranks on the \emph{same common eligible universe} and calculate Spearman's $\rho$ and Kendall's $\tau_b$, with tie-aware handling. Separately, we quantify the mean, median, and maximum absolute displacement of shared publications in their \emph{full scenario-specific rankings}. These two ranking spaces answer different questions and are never substituted for one another.

Discrete annual citation counts can produce extensive score ties, especially for the year-to-year Acceleration difference. We therefore retain the number of tied publications and Top-$K$ boundary-tie diagnostics. Deterministic secondary ordering makes reported Top-$K$ sets reproducible, but does not imply that every membership decision at a tied boundary is substantively unique. We interpret the resulting overlaps as conditional on the documented tie policy.

\subsection{Retrieval context and experimental controls}
Prior to this sensitivity study, two exploratory controlled retrieval exercises used impact-oriented and efficiency-oriented selection strategies. Each involved three successful requests and yielded no additional computable 2024--2025 indicators. These small exploratory observations are reported only as collection context; they do not establish that further requests would be ineffective, nor do they constitute a statistically powered comparison of retrieval policies. In contrast, the v1.8.4 gap sensitivity and v1.8.5 ranking-robustness analyses are strictly read-only and require no external citation-provider requests.

The v1.8.5 output uses a versioned JSON schema (\path{google-scholar-tracker.history-ranking-robustness/v1.8.5}) and records the analysis horizon, scenarios, indicator definitions, tie conventions, eligibility sets, rankings, comparisons, diagnostics, and limitations. Associated CSV and LaTeX tables are generated from the same analytical results. No hypothetical annual observation is written back to the citation database, and no duplicate-review decision is changed. This preserves the distinction between source evidence and counterfactual assumptions and supports independent recalculation from the frozen data and documented code.

\subsection{Scope of inference}
The primary outcomes are descriptive: eligibility counts, Top-$K$ set overlap, common-universe rank correlations, and full-ranking displacement. We do not interpret these metrics as hypothesis tests or as estimates of a population-level effect. The single-profile case study provides a controlled demonstration of a methodological failure mode---the possibility of stable scores but unstable complete rankings---rather than evidence that the same magnitudes occur across disciplines, providers, or researcher populations. Any causal explanation for missing-year omissions remains outside the evidence available in the retained historical records.

\section{Results}
\label{sec:results}
\subsection{Historical coverage and observation uncertainty}
The analysis begins with 616 source records consolidated into 595 publication units. The preceding history audit identifies 246 publications with complete applicable coverage, 61 with internal gaps, 38 with leading gaps, 150 with trailing gaps, and 135 without usable history; these diagnostic categories overlap and therefore do not form a partition. In addition, 34 publications have uncertain publication years. At the observation level, the audit identifies 73 internal, 39 leading, 265 trailing, and 188 no-history missing-year instances. The chronology audit flags 38 possible inconsistency \emph{checks}, which must not be interpreted as 38 confirmed errors or necessarily 38 distinct publications. The gap taxonomy demonstrates that missingness is heterogeneous and that a single global zero-filling rule would conceal meaningful differences in provenance and temporal location.

\subsection{Eligibility sensitivity across missing-data scenarios}
\begin{table}[t]
\centering
\caption{Number of publications for which each indicator is computable under the three scenarios.}
\label{tab:eligibility}
\begin{tabular}{lrrr}
\toprule
Indicator & A & B & C \\
\midrule
Growth & 196 & 257 & 490 \\
Acceleration & 196 & 257 & 490 \\
Revival & 131 & 152 & 177 \\
\bottomrule
\end{tabular}
\end{table}

\begin{table*}[t]
\centering
\caption{Top-$K$ Jaccard overlap between missing-data scenarios.}
\label{tab:overlap}
\begin{tabular}{llrrr}
\toprule
Indicator & Comparison & $J@10$ & $J@20$ & $J@50$ \\
\midrule
Growth & A--B & 1.000 & 1.000 & 1.000 \\
Growth & A--C & 1.000 & 1.000 & 1.000 \\
Growth & B--C & 1.000 & 1.000 & 1.000 \\
Acceleration & A--B & 1.000 & 1.000 & 0.667 \\
Acceleration & A--C & 1.000 & 0.739 & 0.563 \\
Acceleration & B--C & 1.000 & 0.739 & 0.667 \\
Revival & A--B & 0.667 & 0.739 & 0.818 \\
Revival & A--C & 0.667 & 0.739 & 0.818 \\
Revival & B--C & 1.000 & 1.000 & 1.000 \\
\bottomrule
\end{tabular}
\end{table*}

\begin{table*}[t]
\centering
\caption{Rank stability among commonly eligible publications and displacement in complete rankings. All shared-universe Spearman and Kendall correlations are 1.0.}
\label{tab:displacement}
\begin{tabular}{llrrrr}
\toprule
Indicator & Comparison & Shared $n$ & Mean $|\Delta r|$ & Median $|\Delta r|$ & Max $|\Delta r|$ \\
\midrule
Growth & A--B & 196 & 0.88 & 0.00 & 15 \\
Growth & A--C & 196 & 3.67 & 0.00 & 34 \\
Growth & B--C & 257 & 16.62 & 3.00 & 223 \\
Acceleration & A--B & 196 & 37.51 & 45.00 & 61 \\
Acceleration & A--C & 196 & 172.13 & 243.00 & 294 \\
Acceleration & B--C & 257 & 125.27 & 198.00 & 233 \\
Revival & A--B & 131 & 6.50 & 8.00 & 9 \\
Revival & A--C & 131 & 6.50 & 8.00 & 9 \\
Revival & B--C & 152 & 1.33 & 0.00 & 23 \\
\bottomrule
\end{tabular}
\end{table*}

\begin{figure}[t]
\centering\includegraphics[width=.92\linewidth]{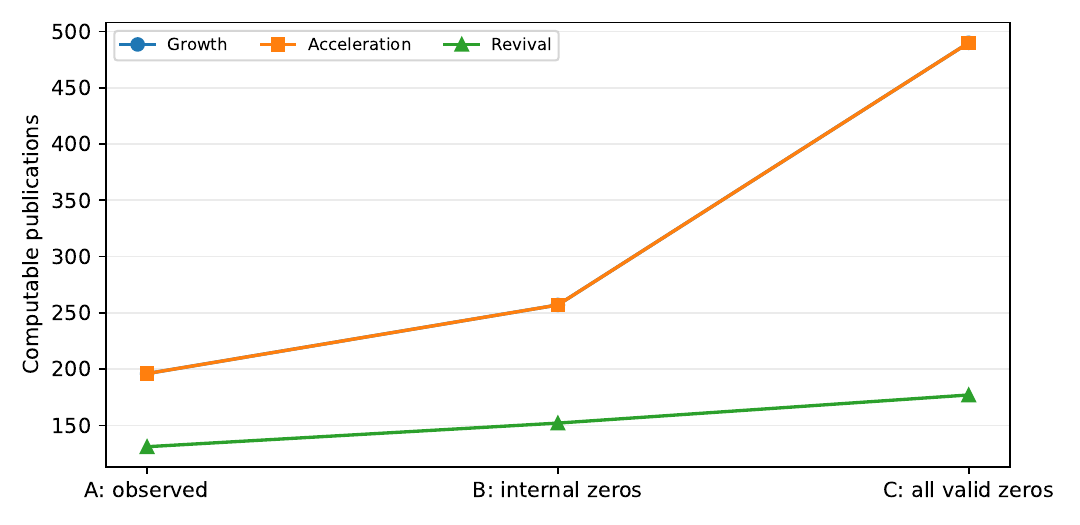}
\caption{Number of publications with computable longitudinal indicators under the three hypothetical missing-data scenarios. Values derive from the v1.8.5 eligibility counts.}
\label{fig:eligibility}
\end{figure}

Figure~\ref{fig:eligibility} shows the eligibility expansion across the nested scenarios. Under the strictly observed-only scenario A, Growth and Acceleration are computable for 196 publications (32.9\% of the 595 consolidated units). Scenario B increases this count to 257 (43.2\%), an absolute gain of 61 publications or 31.1\% relative to A. Scenario C increases it to 490 (82.4\%), a gain of 294 publications or 150.0\% relative to A. Revival is more restrictive: 131 (22.0\%), 152 (25.5\%), and 177 (29.7\%) publications are computable in A, B, and C, respectively. Relative to A, the increases are 21 (16.0\%) and 46 (35.1\%) publications. These denominators describe indicator computability, not the number of papers satisfying a positive-growth or revival-detection threshold.

The sensitivity analysis uses 73 hypothetical zero observations in B and 497 in C, but does not alter stored citation histories. The 497 hypothetical observations are not directly comparable to the sum of audit gap counts: the analysis applies valid publication intervals and indicator-specific observation requirements. Accordingly, the increase in eligibility quantifies dependence on a modeling assumption rather than recovered evidence about actual citation activity.

\subsection{Top-$K$ membership stability}
\begin{figure}[t]
\centering\includegraphics[width=.92\linewidth]{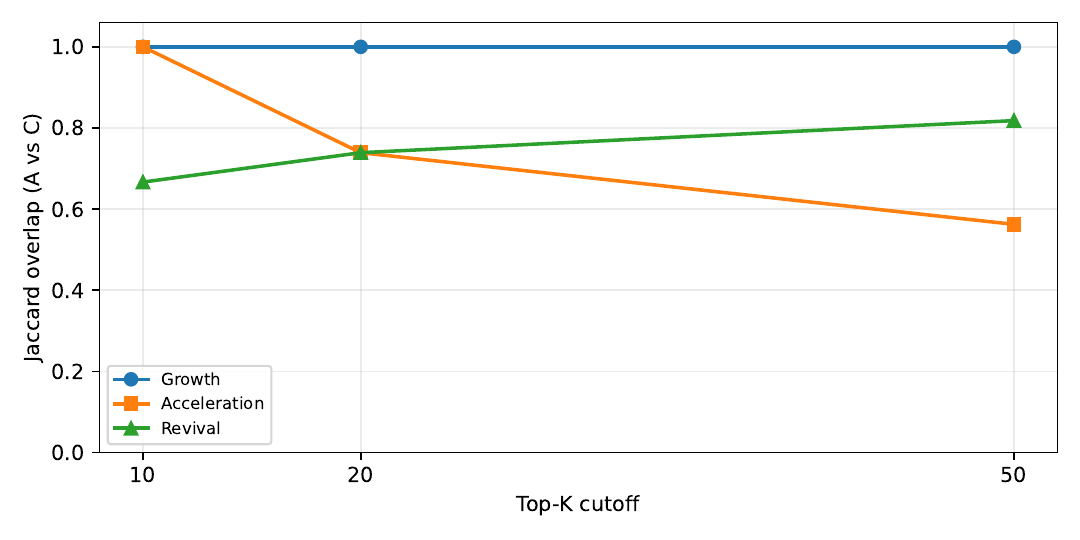}
\caption{Jaccard overlap for the Top-10, Top-20, and Top-50 sets under the strictly observed scenario A and the most permissive scenario C.}
\label{fig:overlap}
\end{figure}

Table~\ref{tab:overlap} reports all three pairwise scenario comparisons at $K\in\{10,20,50\}$. Figure~\ref{fig:overlap} highlights the more demanding A--C comparison. Growth exhibits complete membership agreement ($J@K=1$) at all examined cutoffs, despite the increase from 196 to 490 computable publications. In contrast, Acceleration has perfect Top-10 agreement, but its A--C overlap declines to $J@20=0.739$ (17 shared publications out of 23 in the union) and $J@50=0.5625$ (36 shared publications out of 64 in the union). Thus, 14 publications in the A Top-50 are replaced in the C Top-50. The A--B acceleration Top-50 comparison has $J@50=0.667$, showing that even internal-gap assumptions can affect deeper ranks.

Revival shows a different pattern: its A--B and A--C Top-10 overlap is $0.667$, corresponding to eight shared publications, two entrants, and two exits. At $K=20$, overlap is $0.739$; at $K=50$, it is $0.818$. However, the B--C Revival Top-$K$ sets are identical for all three cutoffs. This demonstrates that additional eligibility in C need not necessarily change the highest-ranked set beyond the changes already introduced by B. The relative sensitivity of an indicator depends on the chosen ranking depth, not merely on the total number of newly eligible publications.

\subsection{Shared-universe score stability and full-ranking displacement}
Table~\ref{tab:displacement} separates fixed-universe correlations from complete-ranking displacement. In every indicator and scenario comparison, Spearman's $\rho$ and Kendall's $\tau_b$ equal 1.0 when calculated on the common eligible universe, consistent with unchanged indicator values for previously computable publications. This perfect conditional agreement should not be read as global ranking invariance. Under A--C, Acceleration exhibits a mean absolute full-ranking displacement of 172.13 positions, a median of 243, and a maximum of 294 across 196 shared publications. The corresponding Growth values are 3.67, 0, and 34, while Revival yields 6.50, 8, and 9 across 131 shared publications.

The large Acceleration displacements arise because the C ranking contains 294 additional eligible publications. A publication can therefore retain the same score and relative position among the original 196 publications but appear far lower in the complete C ranking. Displacement is a measure of the changed ranking universe and its score distribution, not evidence that previously observed annual citations or computed scores were modified.

\subsection{Ties and sensitivity at Top-$K$ boundaries}
The output reveals substantial score ties. Under A, 185 of 196 Acceleration publications (94.4\%) belong to tied-score groups; under C, 479 of 490 (97.8\%) do. Growth also has extensive ties (174 of 196 under A and 468 of 490 under C), whereas Revival has 75 of 131 and 121 of 177, respectively. The implementation uses competition ranks for displayed score ranks and a deterministic secondary order for Top-$K$ membership, while correlations employ average ranks on the common universe.

Some Top-$K$ boundaries cut through a tie group. For example, in the A--C Acceleration Top-20 comparison, the A boundary has a tied score shared by five publications, of which only three are included. This means that membership-based overlap partly reflects the stated tie-breaking policy. Accordingly, the numerical Jaccard results should be interpreted as reproducible \emph{policy-conditional} outcomes; a tie-inclusive or tie-averaged sensitivity analysis would be a useful complementary experiment before making broader claims about the stability of tied rankings.

\subsection{Summary of findings relative to the research questions}
For RQ1 (coverage and provenance), the audit documents multiple kinds of missingness and uncertain publication chronology; the retained data cannot establish the semantics of omitted years. For RQ2 (eligibility sensitivity), the number of computable publications increases from 196 to 490 for Growth and Acceleration and from 131 to 177 for Revival between scenarios A and C. For RQ3 (ranking robustness), indicator values and shared-universe ordering remain unchanged, yet complete-ranking positions can shift as the eligible set expands. Top-$K$ robustness is indicator- and depth-dependent: Growth is stable at all tested cutoffs, Acceleration becomes less stable for deeper lists, and Revival is sensitive already at the Top-10 level. All findings are descriptive for the studied portfolio and should not be interpreted as evidence that hypothetical zeros reflect the provider's true omission semantics.

\section{Discussion}
\label{sec:discussion}

\subsection{Eligibility is a separate dimension of bibliometric reliability}
The central methodological implication is that missing-data handling changes not only numerical estimates but also the population to which an indicator can be applied. In the strict observed-only scenario, 196 publications have computable Growth and Acceleration scores, whereas 490 become computable when all valid missing years are hypothetically replaced by zeros. The difference is 294 publications, or a 150\% increase relative to the observed-only baseline. For Revival, the corresponding change is from 131 to 177 publications, an increase of approximately 35.1\%. These changes should not be interpreted as an improvement in the underlying data: they arise from a stronger assumption about unobserved years. Reporting only a final ranking without its eligibility denominator would obscure this dependence.

This distinction has implications for responsible research assessment. A publication absent from a longitudinal ranking may lack the necessary historical observations rather than exhibit weak citation performance. Conversely, inclusion under an imputation policy does not demonstrate that the missing citation counts were truly zero. Consequently, ranking reports should state the number of candidate publications, the number eligible for each indicator, the observation requirements, and the assumptions applied to unavailable years. Such reporting complements the transparency principles advocated for bibliometric indicators \citep{hicks2015leiden,waltman2016review}.

\subsection{Why perfect rank correlations do not establish robust rankings}
The observed Spearman and Kendall correlations of 1.0 on the common eligible universe are compatible with substantial changes in full-ranking positions. The reason is structural: previously eligible publications retain their scores and relative ordering, while newly eligible publications are inserted into the ranked population. In the A--C Acceleration comparison, the 196 common publications have a mean absolute full-ranking displacement of 172.13 positions and a maximum displacement of 294. Such changes do not contradict perfect common-universe correlation; the two statistics describe different objects.

A useful reporting framework therefore separates four questions: (i) which publications can be scored; (ii) whether a publication's score changes; (iii) whether its order relative to other shared publications changes; and (iv) whether it remains in a decision-relevant Top-$K$ set. A correlation computed after restricting the comparison to shared publications cannot answer the fourth question. Conversely, a Top-$K$ membership change alone does not prove that the score of any previously eligible publication was modified.

\subsection{Indicator-specific sensitivity and the role of ties}
The three indicators respond differently to the hypothetical-zero scenarios. Growth retains identical Top-10, Top-20, and Top-50 membership between A and C, whereas Acceleration has Jaccard@50 approximately 0.563 and Revival has Jaccard@10 approximately 0.667. These observations do not support a general claim that one indicator is inherently more robust: they are conditional on the studied citation distributions, observation window, indicator definitions, and tie policy.

Ties warrant particular caution. In the observed-only scenario, 185 of 196 Acceleration publications belong to tied-score groups; under scenario C, 479 of 490 do. A deterministic secondary ordering makes the exported Top-$K$ lists reproducible, but it does not eliminate the analytical arbitrariness of selecting only part of a tied group at a cutoff. A tie-inclusive Top-$K$ analysis, or a range of overlaps across admissible tie resolutions, would be a valuable complementary experiment. Until then, membership-overlap estimates should be described as conditional on the documented tie-breaking procedure.

\subsection{Missingness, provenance, and the limits of additional retrieval}
A missing annual observation has multiple plausible explanations: an unreported zero, incomplete historical coverage, provider-side revision, or a collection or parsing limitation. The available parsed histories do not identify which explanation applies to a particular omitted year. Thus, scenario A provides the conservative evidentiary baseline, while scenarios B and C quantify sensitivity to increasingly permissive assumptions rather than supply reconstructed ground truth.

The distinction also guides data acquisition. Controlled retrieval experiments in the tracker yielded no additional usable annual observations in their small samples. This observation motivates explicit accounting of query costs and information gains, but it does not establish that further retrieval would never be useful. Prospective capture of raw provider responses, request outcomes, timestamps, and parser versions would enable stronger future tests of omission semantics. Chronological anomalies and unresolved duplicate relationships should remain visible as provenance warnings instead of being silently corrected.

\subsection{Software and reusable skill as a research contribution}
Beyond the numerical case study, the work is supported by a reusable computational implementation, referred to here as the \emph{Google Scholar Tracker skill}. The skill operationalizes a workflow that includes citation-data collection, local snapshots, duplicate-record review and consolidation, historical-coverage auditing, indicator computation, hypothetical missing-data scenarios, and ranking-robustness reporting. Its command-line interface and machine-readable exports are intended to make the methodological decisions explicit and repeatable. The scientific contribution lies not merely in automating data retrieval, but in coupling each reported result to its eligibility rules, observation provenance, and sensitivity assumptions.

The software is publicly available as version 1.8.5 under the MIT License at \url{https://github.com/jppbsi/google-scholar-tracker}, with an immutable software archive on Zenodo \citep{papa2026googlescholartracker}. The repository contains source code, a command-line interface, operational skill instructions, and automated tests. The public distribution excludes the local research database and other private study artifacts; consequently, access to the code alone does not guarantee exact regeneration of every numerical result reported here.

The archived version-specific DOI identifies the software release used in this study, rather than a mutable development branch. The repository also provides citation metadata. Full independent numerical replication additionally requires the frozen input observations, experimental configuration, and any data-sharing permissions; the public code release does not by itself supply the private citation database.

\subsection{Implications for future evaluation and generalization}
The framework is applicable in principle to other portfolios and annual-count providers, provided their temporal semantics and data-access conditions are explicitly characterized. However, the present evidence comes from a single researcher profile. Replication across career stages, disciplines, and citation-density regimes is needed to assess how frequently the observed patterns recur. In addition, comparison with an independently validated annual-citation source could help distinguish missing observations from actual zero-citation years. These extensions would turn the current case study into a broader evaluation of bibliometric data reliability rather than an unsupported generalization from one portfolio.

\section{Threats to Validity}
\label{sec:validity}

\subsection{Construct validity: what is actually measured?}
Our analysis concerns the \emph{availability and ranking consequences of reported annual citation observations}, not the true annual citation trajectories of the underlying publications. The historical tables returned by the provider are treated as year-specific counts; an absent row is not independently verified to represent zero citations. Scenarios B and C are therefore controlled counterfactual assumptions, not estimates of latent ground truth. Conflating the two would overstate the evidence and could lead to unwarranted claims about citation growth, acceleration, or revival. The chosen indicators are operational definitions of longitudinal activity, not comprehensive measures of research quality or impact. In particular, the Revival score depends on the historical reference window, age eligibility, and treatment of a zero baseline; these choices constrain construct validity and should not be interpreted as a universal definition of scholarly resurgence.

\subsection{Internal validity: metadata, chronology, and consolidation}
The starting records are not necessarily independent publications. Bibliographic variants, preprints, and versioned entries may refer to the same scholarly output. Consolidation reduces this risk but introduces dependence on duplicate-component review, record selection, and the assignment of a historical reference. Residual false merges or missed duplicates can affect denominators, annual histories, and rank membership. The audit also identified 38 possible chronology inconsistencies, including annual observations preceding recorded publication years. These are \emph{flags}, not 38 verified errors: online-first publication, metadata discrepancies, and provider revisions are plausible explanations. We retained such uncertainty rather than silently rewriting publication years. A future adjudication study should record independent reviewer decisions and measure how alternative consolidation and chronology policies change the reported results.

\subsection{Measurement and data-source validity}
Google Scholar coverage, bibliographic metadata, and citation counts can change over time. SerpApi acts as an access layer and may introduce its own response format, omissions, or parsing constraints. Because complete historical raw responses were not prospectively retained, the present data cannot identify whether a missing year originates in the source, the intermediary, or the collection pipeline. Moreover, an annual series reconstructed from later queries need not match the information that would have been available at the corresponding historical date. Snapshot timestamps and versioned parsers mitigate some reproducibility risks but cannot reconstruct unobserved provider states. Validation against independently retained raw responses and, where feasible, a second citation source is necessary before assigning semantic meaning to omissions.

\subsection{Analytical validity: scenario design and indicator definitions}
The three scenarios span explicit policies for missing annual observations but do not exhaust the space of plausible missing-data mechanisms. In particular, missingness may be associated with publication age, citation volume, or provider coverage, so it need not be missing completely at random. Scenario C is intentionally permissive; it should not be interpreted as a recommended imputation policy. The analysis uses a fixed 2024--2025 evaluation window and specific Growth, Acceleration, and Revival implementations. Other windows, baselines, minimum ages, or threshold choices may change both eligibility and rankings. Sensitivity conclusions are therefore conditional on the defined experimental configuration rather than intrinsic properties of the indicators.

\subsection{Ranking validity: ties, denominators, and comparison universes}
Ranking statistics answer different questions. Spearman and Kendall coefficients computed only for publications eligible under both scenarios quantify relative ordering \emph{within the common universe}; they do not measure whether newly eligible publications displace those records in a complete ranking. Conversely, Top-$K$ Jaccard overlap depends on the cutoff and on a deterministic secondary ordering for tied scores. This dependence is especially material for Acceleration, where tied-score groups are extensive. A cutoff that intersects a tie group may produce different membership under another defensible tie policy, even when the primary scores are unchanged. We therefore interpret overlap and displacement as conditional descriptive statistics. Future analyses should include tie-inclusive cutoffs, uncertainty bounds over admissible tie resolutions, and several evaluation windows.

\subsection{External validity and statistical conclusion validity}
The case study uses 595 consolidated publications from a single researcher profile. This portfolio may differ from other researchers in publication age, field composition, collaboration patterns, and citation density. Neither the magnitude of eligibility expansion nor the observed stability of Growth Top-$K$ should be generalized to a population of researchers. The comparisons are paired deterministic sensitivity analyses, not independent randomized samples; no population-level statistical significance follows from their numerical differences. Similarly, the exploratory retrieval experiment used only three requests per strategy and yielded no additional usable annual observations in those trials. That sample cannot establish the relative effectiveness or general futility of additional retrieval. Multi-profile replication with prespecified inclusion rules and uncertainty summaries is needed for broader inference.

\subsection{Reproducibility, implementation, and release limitations}
The software workflow is implemented in the publicly released Google Scholar Tracker skill and Python package (version 1.8.5), archived under DOI 10.5281/zenodo.24170285. However, publication of the source code does not make the private citation database or frozen experimental outputs publicly available. Exact replication may also depend on third-party API access, changing provider data, and restrictions on redistribution. Although a pre-release credential review and automated tests were performed, future independent security and dependency audits, pinned environments, and broader test coverage remain valuable. A synthetic fixture and frozen machine-readable outputs would allow verification of the analysis without requiring access to private credentials or mutable external services.

\section{Conclusion and Future Work}
\label{sec:conclusion}
Incomplete annual citation histories create a problem that cannot be reduced to selecting a numerical imputation rule. They determine \emph{which publications can be evaluated}, which scores can be computed, and which publications enter decision-relevant rankings. This study introduced a provenance-aware, read-only framework that distinguishes observed annual counts from absent observations, audits temporal coverage, evaluates three explicitly labeled missing-data scenarios, and measures ranking sensitivity using eligibility counts, Top-$K$ overlap, common-universe correlations, and complete-ranking displacement.

The case study comprised 595 consolidated publications and a 2024--2025 analysis window. Under the strict observed-only scenario, Growth and Acceleration were computable for 196 publications; under hypothetical zero assignment for all valid missing years, 490 became computable. Revival eligibility increased from 131 to 177. Despite unchanged relative ordering among publications common to the compared scenarios, full-ranking outcomes were not uniformly stable: Growth retained identical Top-10, Top-20, and Top-50 membership between A and C, whereas Acceleration reached Jaccard@50 of approximately 0.563 and Revival reached Jaccard@10 of approximately 0.667. For Acceleration, mean absolute complete-ranking displacement reached 172.13 positions. These results show that a perfect correlation on the shared eligible set is insufficient evidence of overall ranking robustness.

The principal methodological contribution is a reproducible separation of \emph{observation provenance}, \emph{indicator computability}, and \emph{ranking robustness}. The accompanying computational contribution is the Google Scholar Tracker skill, which implements the workflow through auditable data collection, consolidation, historical-coverage checks, indicator computation, scenario analysis, and machine-readable exports. The implementation is publicly released on GitHub and permanently archived as version 1.8.5 on Zenodo \citep{papa2026googlescholartracker}. This makes the software inspectable while leaving the underlying private observational dataset subject to separate access and redistribution constraints.

The findings also suggest a practical reporting standard for longitudinal bibliometrics: every indicator-based ranking should disclose its candidate population, eligible denominator, required annual observations, missing-data policy, tie-breaking rule, evaluation window, and sensitivity to alternative admissible assumptions. When the meaning of an omitted year has not been verified, an observed-only analysis should remain the evidentiary baseline, with hypothetical imputations reported separately as sensitivity experiments rather than substituted for observed data.

Future work will proceed along four complementary directions. First, prospectively captured raw responses and independently checked citation histories will be used to investigate the semantics of omitted years and temporal revisions. Second, manual review of chronology flags and duplicate components will quantify the effect of alternative metadata decisions. Third, tie-inclusive ranking comparisons, additional time windows, and alternative indicator definitions will test the stability of the current conclusions. Fourth, multi-researcher, multi-discipline replications will examine generalizability. The archived public software release provides a foundation for independent evaluation and extension; a synthetic reproducibility dataset and fully pinned environment remain useful future additions. These steps are necessary to move from a carefully controlled case study toward broader evidence about the reliability of longitudinal bibliometric rankings.

\section*{Reproducibility Statement}
The reported numerical tables were generated from the v1.8.5 JSON supplied with the study, using the 2024--2025 window and $K=10,20,50$. The v1.8.4 gap statistics were taken from the corresponding audit and sensitivity reports. The source code of the Google Scholar Tracker framework, including the implementation of the proposed bibliometric indicators and ranking robustness analyses, is publicly available at \url{https://github.com/jppbsi/google-scholar-tracker}. Version 1.8.5 of the software has been permanently archived on Zenodo under DOI \href{https://doi.org/10.5281/zenodo.24170285}{10.5281/zenodo.24170285} \citep{papa2026googlescholartracker}. The archived release provides a persistent reference to the implementation used in this study, supporting reproducibility and further methodological development. Exact reproduction of the reported numerical results additionally requires the frozen study data and configuration, which are not included in the public code repository; data sharing remains subject to provider terms and researcher privacy.


\end{document}